\documentclass[12pt, onecolumn,draftcls]{IEEEtran}
\usepackage{enumerate}
\usepackage{amsmath,amssymb,bm,amsfonts}
\usepackage{graphics,graphicx,epsfig,subfigure}
\usepackage{algorithm,algorithmic}
\usepackage{theorem}
\usepackage{threeparttable}
\usepackage{stfloats}
\usepackage{cite}
\usepackage{multirow}
\usepackage{mathrsfs}
\usepackage{color}
\usepackage{multirow}
\usepackage{subeqnarray}
\usepackage{cases}
\usepackage{colortbl}
\usepackage{wasysym}
\usepackage[dvipsnames]{xcolor}
\usepackage{url}
\usepackage{hyperref}
\usepackage{epstopdf}

\newtheorem{mytheorem}{Theorem}
\newtheorem{mylemma}{Lemma}
\newtheorem{myproof}{Proof}
\newtheorem{mydef}{Definition}
\begin{document}
\newcommand*{\QEDA}{\hfill\ensuremath{\blacksquare}}
\def\arrow{{\rightarrow}}
\def\N{{\mathcal{N}}}
\def\B{{\mathcal{B}}}
\def\E{{\mathcal{E}}}
\def\I{{\mathcal{I}}}
\def\diag{{\textrm{diag}}}
\def\i {{ -i}}
\def\ci{\perp\!\!\!\perp} % from Wikipedia
\newcommand\independent{\protect\mathpalette{\protect\independenT}{\perp}} % symbols-a4, p.106
\def\independenT#1#2{\mathrel{\rlap{$#1#2$}\mkern2mu{#1#2}}}

\title{When Renewable Energy Meets Building Thermal Mass: A Real-time Load Management Scheme}
\author{Yan Shen,  Zhonghao Sun, and Qinglong Wang

%\thanks{The authors are with the Department of Electrical and Electronic Engineering, The University of Hong Kong, Pokfulam Road, Hong Kong (e-mail: \{duinan, ycwu\}@eee.hku.hk).}
%%
% <-this % stops a space
%%%\thanks{H. Vincent Poor is with the Department of Electrical Engineering, Princeton University, Princeton, NJ 08544 USA (email: poor@princeton.edu).}
}
\maketitle

\IEEEpeerreviewmaketitle
\begin{abstract}
We consider the optimal power management in
renewable driven smart building MicroGrid under noise corrupted
conditions as a stochastic optimization problem. We
first propose our user satisfaction and electricity consumption
balanced (USECB) profit model as the objective for optimal power
management. We then cast the problem in noise corrupted
conditions into the class of expectation maximizing in stochastic
optimization problem with convex constraints. For this task, we
design a Bregemen projection based mirror decent algorithm
as an approximation solution to our stochastic optimization
problem. Convergence and upper-bound of our algorithm with
proof are also provided in our paper. We then conduct a broad
type of experiment in our simulation to test the justification of
our model as well as the effectiveness of our algorithm.
\end{abstract}

%
%%%%%%%%%%%%%%%%%%%%%%%%%%%%%%
% Introduction
%
%%%%%%%%%%%%%%%%%%%%%%%%%%%%%
\section{Introduction}
The concept of smart grid that incorporating power transmission
system and end users in an integral system has been
gradually embraced by independent system operators(ISO) \cite{wang2016smart}.
Smart buildings utilizes the advancement of power electronics,
communication technology \cite{cai2010cfo} and control theory to allow for
the dynamic interaction between power grid and end users
without human intervention. The drivers for smart grid are
not only the financial benefit for end users generated by
energy consumption cost reduction but also the positive effect
on grid operation stability. On the other perspective, smart
building is considered as an ideal active participant of demand
response program for its ability to regulate power consumption
using its thermal mass capacities \cite{braun2003load}. Deployed with advanced
load control technology, smart building can smartly coordinate
their power consumption with smart grid’s load condition
\cite{sinopoli2009smart}. However smart building’s power consumption pattern that
copeaks with power grid’s load occupancy condition limits
its load regulating ability. Power system’s stability condition
will become fragile when building’s air-conditioning power
suddenly increases in the condition of high volume load in
power grid.

The northeast blackout of 2003 can be provided as a
typical example of cascaded widespread power system failure
triggered by surging building air conditioning loads in a hot
summer. According to the technical analysis conducted by
North American Electric Reliability Council \cite{council2004technical}, there is no
matter that the stringent power consumption caused by high
air-conditioning usage served as a catalyst to worsen the whole
event, though the direct cause that triggers the power failure
is the failing of the alarm and logging system in FirstEnergy
control room \cite{bolognani2013distributed}.

This motivates us to find a substitutional approach to mitigate
the impact of smart building’s overloading in peak hour.
Thanks to the deregulation of electricity market and the advancement
of MiroGrid technology, supplying smart building
with local renewable generation through the interconnection
of MicroGrid becomes a feasible solution to reduce smart
building’s reliance on Grid. As renewable generation copeaks
with the smart building’s power consumption (typically in
a hot summer afternoon), local renewable can be efficiently
utilized by smart buildings nearby without the need of long
distance transmission. In this paper, we discuss the possibility
of incorporating the smart building load and renewable
generation together in a MicroGrid by studying the stability
condition of power grid under load and generation fluctuation.
Power management strategy is deployed in our MicroGrid to
achieve the dual goal of minimizing outside power consumption
and maximizing user satisfaction. Measurement error and
renewable generation uncertainties are also considered in our
paper.

\indent The {\bf main contribution} of this paper is three folds: In one perspective, we propose the concept of supplying smart building with local renewable energy as a complimentary power source to commercial grid. In another perspective, we develop our user satisfaction and electricity consumption balanced (USECB) profit model as a objective goal for power management strategy in MicroGrid. Our contribution is not only limited in system model but also in theory of algorithms, we design a Bregemen projection \cite{du2018proactive} based mirror decent algorithm as a stochastic optimization scheme with convergence analysis.\\
%\indent\textbullet First, we propose the concept of supplying smart building with local renewable energy as a complimentary power source to commercial grid. Smart building is connected to MicroGrid in which the renewable generation is incorporated. Power from outside grid can flow to smart building as MicroGrid is connected to commercial grid.\\
%\indent\textbullet Second, we develop our user satisfaction and electricity consumption balanced(USECB) profit model as a objective goal for power management strategy in MicroGrid. User's satisfactory and MicroGrid's stable condition are both considered in our model. In our USECB model, user's satisfaction and increasing local renewable dependency are considered in a balanced way.\\
%\indent\textbullet Finally, we design a Bregemen projection based mirror decent algorithm as a stochastic optimization scheme. This algorithm is extraordinary effective in a noise contaminated environment. Convergence bound is also theoretically proved in our paper.\\
\indent Our paper is organized as follows: In section(\ref{system model2}), we first introduce the back ground of power transmission network and applies the linear approximation model  in our MicroGrid as system model. In section(\ref{problem formulation}), we formulate our optimal power management problem in renewable driven smart building MicroGrid. In section(\ref{stochastic solver}), we introduce our Bregemen Projection based mirror decent algorithm. In section(\ref{simulation}), we show the performance of our model and algorithm via simulation.
\cite{congestionimpact}

%For DSRC communication, $10$ MHz of spectrum in the $5.9$ GHz band is used.

% \usepackage{multirow}
%\begin{table*}
%\center
%\caption{Differences between Vehicular CPS and Cellular Networks}
%\label{Tab1}
%\begin{tabular}{cll}
%\hline
%\rowcolor[HTML]{EFEFEF}
%&\quad \quad \quad \quad \quad Vehicular CPS   &\quad \quad \quad  \quad \quad  \quad Cellular Networks   \\
%\hline
%&High moving speed resulting in large CFO       &
%\quad Relative low moving speed and small CFO \\ \hline
%&Information is broadcasted to all neighboring vehicles
%&\quad Communication between one pair of cellular and base station \\\hline
%&Highly dynamic network topology &
%\quad Static network topology \\\hline
%& Distributed & \quad Centralized control and computation by cellular\\
%\hline
%\end{tabular}
%\end{table*}

\section{Related Work}
Up to now, many studies have proposed the concept of smart building's load control using its thermal mass capacities. P.Xu has studied smart building's load control potential in Public Interest Energy Research (PIER) Program that was conducted at the Santa Rosa Fedral Building. Field test were performed in two buildings \cite{xu2005demand}. The result of load monitoring together with comfort surveys concludes the potential to perform demand response in commercial buildings while maintaining acceptable comfort conditions. However these studies are performed in the framework of passively following grid's regulation signal while omitting the part of interactive influence by considering the actual power grid model. B.Ramanathan has incorporated the interactive influence part in power grid in their load control modeling. However they only use an empirical one from Nordic power system for a specific performance without universal applicability \cite{Gu, ramanathan2008framework, henze2004evaluation}.\\
\indent As generally studying the impact of smart building's load control in power grid need to adopt a concise and exact power grid model, DistFlow model \cite{farivar2013branch} is among the most popular exact power grid model. However DistFlow model can not be directly applied because of the inconvex nature of the model. S.Low has developed a convex relaxation of DistFlow model \cite{Wang, farivar2012branch}. This model is effective in solving most of the optimal power flow (OPF) problem. Though the theoretical exactness is only proved in radical cases, simulation evidence has shown that its exactness can be broadly extended to general cases. S.Bolognani has generalized one DC power flow model to AC power grid systems. This model is developed by the first order Tyler approximation of power injection equations. High exactness for MicroGrid in grid connected case has been tested in simulations. 

 The concept and enabling technologies for the Customer-driven MicroGrid has been extensively studied in many researches. S.Suryanarayanan has surveyed the enabling technologies of the Customer-driven MicroGrid \cite{suryanarayanan2009enabling}. P. Dondi has reviewed the position of distributed generation with respect to the installation and interconnection of such units \cite{dondi2002network}. Actually many of these technologies have been applied to the MicroGrid testing systems. The concept of MicroGrid has gone beyond its seminar stage \cite{farhangi2010path}. Actual onsite testbed has been installed as a precursor for later residential product. Consortium for Electric Reliability solution(CERTS) testbed has been experimented in Lawrence Berkeley lab campus \cite{lasseter2011certs}. Distributed renewable generation resources \cite{asilomar} have been integrated in CERTS MicroGrid. Testing on MicroGrid components has also been extensively conducted by ISET in Germany. The More Project located at Bronsbergen
Holiday Park, located near Zutphen in the Netherlands aims at the increasing penetration of microgeneration in electrical networks through the exploitation and extension of MicroGrid concept \cite{katiraei2008microgrids}.

%%%%%%%%%%%%%%%%%%%%%%%%%%%%
% Motivating Example
%
%%%%%%%%%%%%%%%%%%%%%%%%%%%%%%
\section{Supplying Smart Building with Local Renewables:Motivations and Challenges}\label{Motivation}
\begin{figure}[b]
\centering
\includegraphics[width=3.5in]{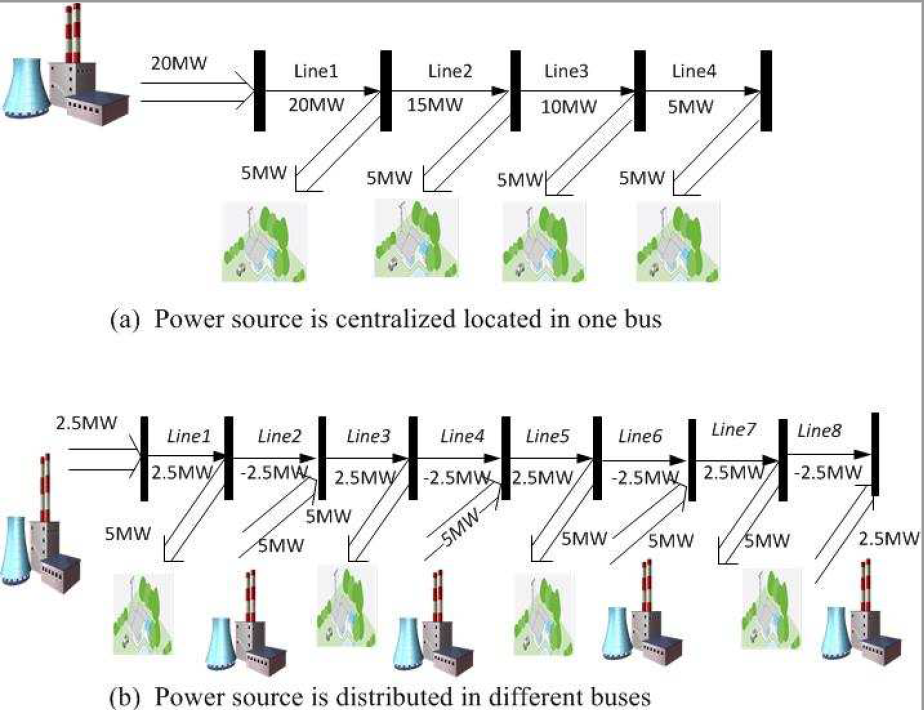}
\caption{Power load on the transmission line is different as a result of different power source placement.}
\label{motiv_exa}
\end{figure}

Despite of smart building's load control ability by using its thermal mass capacity, researches are mainly focused on adjusting its power consumption pattern to adapt to the peak and valley signals generated by conventional grid without noticing the fact that the peak and valley signals are generating by the grid according to the congestion conditions of its major transmission branch rather than the ability of its instantaneous generating capacities to meet its demands. As the distribution network of conventional commercial grid is typically in a radical structure, the transmission of electricity power from generation unit to terminal customs are competing for some of the major transmission branches. The commercial grid has to restrict the load in some of consumption buses to avoid overloading its transmission system when heavy volumes of the electricity power are transmitting from the centralized generation unit to remotely distributed customs \cite{informationmatrix} simultaneously, even if the grid has excessive generating capacities. This phenomenon becomes more obvious when renewable are incorporated in the generation, as the solar energy copeaks with smart building demands.\\
\indent We provide the following example to illustrate the advantage of supplying smart building with local renewable generations. In Figure(\ref{motiv_exa})a, a 20MW capacity generation unit is centrally located in bus 0. The consumptions are equally distributed in the remaining buses. All the buses are connected in a tree topology. We assume in this example that all the consumption buses has the same power load of 5MW. Therefore there is a 20MW power flow on the transmission line 1, while the power flow on transmission line 4 is 5MW. As the degree of loading caused thermal loss on the transmission line is proportional to the square of power that flow on the transmission line, the load on transmission line 1 is 15 times heavier than the load on transmission line 4 which will become the bottleneck for the whole transmission system.\\
\indent However in Figure(\ref{motiv_exa})b, a total capacity of 20MW generation unit is distributed in bus 0, 2, 4, 6 and 8. Bus 0, 2 ,4, 6 and 8has the corresponding generation capacity of 2.5MW, 5MW , 5MW, 5MW and 2.5MW. The power load at the consumption buses are all 5MW which is the same as the previous condition. Under the conditions in our second example, the power flow in every transmission line is 2.5MW. All the transmission lines in this power system has the same load. \\
\indent From the previous examples, we observe that the overloading of the main transmission branch will not become the bottleneck for renewable energy utilization when the renewable energies are mainly transmitting in its local branch. This motivate us to study the feasibility of integrating renewable generation and smart building in the same local MicroGrid and to further design a power management strategy to stabilize MicroGrid's operation condition thus utilizing the renewable energy in most efficient way. The MicroGrid we are studying has the following characteristics:\\
\indent The power load in smart buildings and geo-distributed \cite{duJMLR} renewable generation sites are connected to the MicroGrid. The power load for a single smart building is connected to one transformer. The renewable power generating units in one single location such as PV panel modules and wind turbine farms is also connected to one transformer. Each transformer is considered as one bus in the MicroGrid. Different buses in the microgrid are connected by the transmission lines. The energy generated by the renewable generating site is transmitting through the lines to smart building users. Contrary to the case of centralized generation, in which the electrical power flows in one way follows the direction from the generation bus as root to its sub-node, the actual direction of power flow is dynamic and determined by the realtime network operational condition. Finally, the MicroGrid requires or provides additional power from the grid. Microgrid connects to the grid through point of common coupling(PCC) through which the additional power can flow into the MicroGrid when renewable generation in MicroGrid is insufficient. Power generating buses have an inflexible and unpredictable power influx. Power consumption buses that are connected to the smart building has a controllable loads. Figure 1 draws an abstract picture of our renewable driven smart building power MicroGrid.

\indent Realizing the concept of integration of distributed generation and smart buildings in an autonomy MicroGrid and further operating it properly are non-trivial tasks. Towards accomplishing these objectives, we need to address the following challenges:\\
%\textbullet {\bf Challenge 1:} \textit{Protecting Power System Stability}: The reliable operation of electrical power system depends on keeping the system within the safe operation limits. Power system parameters such as power flows and bus voltages must not violate its safe operation standard in order to avoid severe failure. As these power system parameters on each buses are determined by KCL,KVL and Ohm's laws once power load on each bus is given, failing to regulate the load on buses may seriously affect system's stability by disturbing power system's parameter too much to violate their safety constraints. Power system becomes even more volatile in the distributed power generation cases when power flow and reverse power flow coexists under frequent fluctuations.\\
\textbullet {\bf Challenge 1:} \textit{Protecting Power System Stability}: The reliable operation of electrical power system depends on keeping the power system parameter within the safe operation limits. These parameters such as power flows and bus voltages are determined by KCL,KVL and Ohm's laws once power load on each bus is given. Failing to regulate the load on buses may seriously affect system's stability. Power system becomes even more volatile when distributed power generation is included.\\
\textbullet {\bf Challenge 2:} \textit{User satisfaction related load}: The thermal dynamics of the smart building relates to its air conditioning power consumptions. As users may feel uncomfortable when indoor temperature deviates the set point too much. Frequently regulating the smart building's air conditioning load may compromise user's satisfaction.\\
\textbullet {\bf Challenge 3:} \textit{System errors in data acquisition}: As converting these data from its original state to state in digital circuit in a high sampling frequency need complex structures such as A/D and amplifiers, introducing white Guassian noise in the data collected from the sensors is inevitable. We should not sacrifice our accumulative performance too much under unreliable and random data.\\
\indent To this end, we handle the previous stated difficulties by modeling the power management scheme in renewable driven smart building MicroGrid as convex optimization problem with a competitive stochastic solver. We will give an approximative power system model in section (\ref{system model2}), a user satisfaction model in section(\ref{problem formulation}) and proposed a stochastic iterative algorithm in section (\ref{stochastic solver})
\begin{figure}
\centering
\includegraphics[width=3.5in]{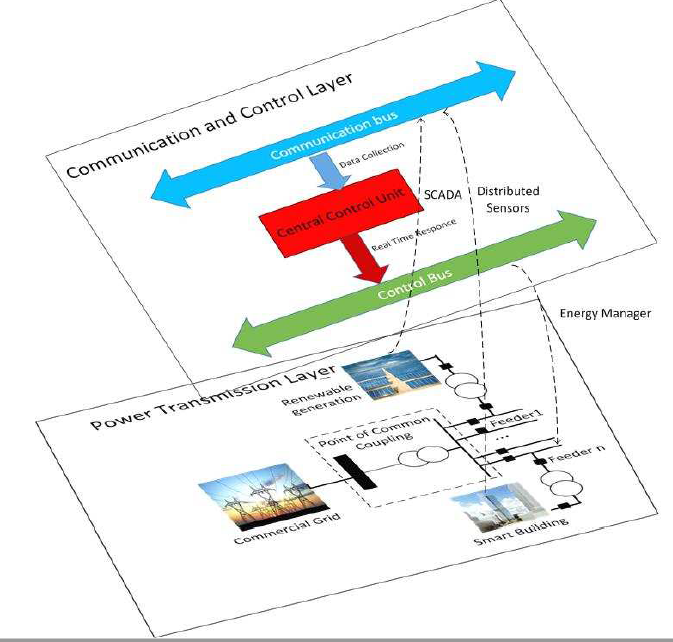}
\caption{System architecture of our renewable driven smart building power MicroGrid}
\label{system model}
\end{figure}
%\begin{figure}[ht]
%  \centering
%\mbox{\subfigure[Compensating the Doppler shift $f_{A,B}$ and  and  $f_{A,C}$ cannot compensate the Doppler shift between B and C.]{\epsfig{figure=./figures/eg1,width=1.8in}}\label{eg1}
%}
%\qquad  \qquad
%\mbox{
%      \subfigure[Compensating different Doppler shift at each vehicle to make there is no relative Doppler shift between each pair of communication link.]{\epsfig{figure=./figures/eg2,width=1.8in}}\label{eg2} }
%\caption{Compensating different Doppler shift at each vehicle to make there is no relative Doppler shift between each pair of communication link.}
%  \label{fig:SubF}
%\end{figure}
%
%\begin{figure}[!ht]
%  \centering
%{\epsfig{figure=./Figures/eg1,width=1.5in}}
%\caption{Compensating the Doppler shift $f_{A,B}$ and  and  $f_{A,C}$ cannot compensate the Doppler shift between B and C.}
%\label{MSE-Iter}
%\end{figure}
%\begin{figure}[!ht]
%  \centering
%{\epsfig{figure=./Figures/eg2,width=1.5in}}
%\caption{Compensating different Doppler shift at each vehicle to make there is no relative Doppler shift between each pair of communication link.}
%\label{MSE-Iter}
%\end{figure}

\section{System Model
}\label{system model2}

\subsection{General Power Grid Model}
\begin{figure}[b]
\centering
\includegraphics[width=3.0in]{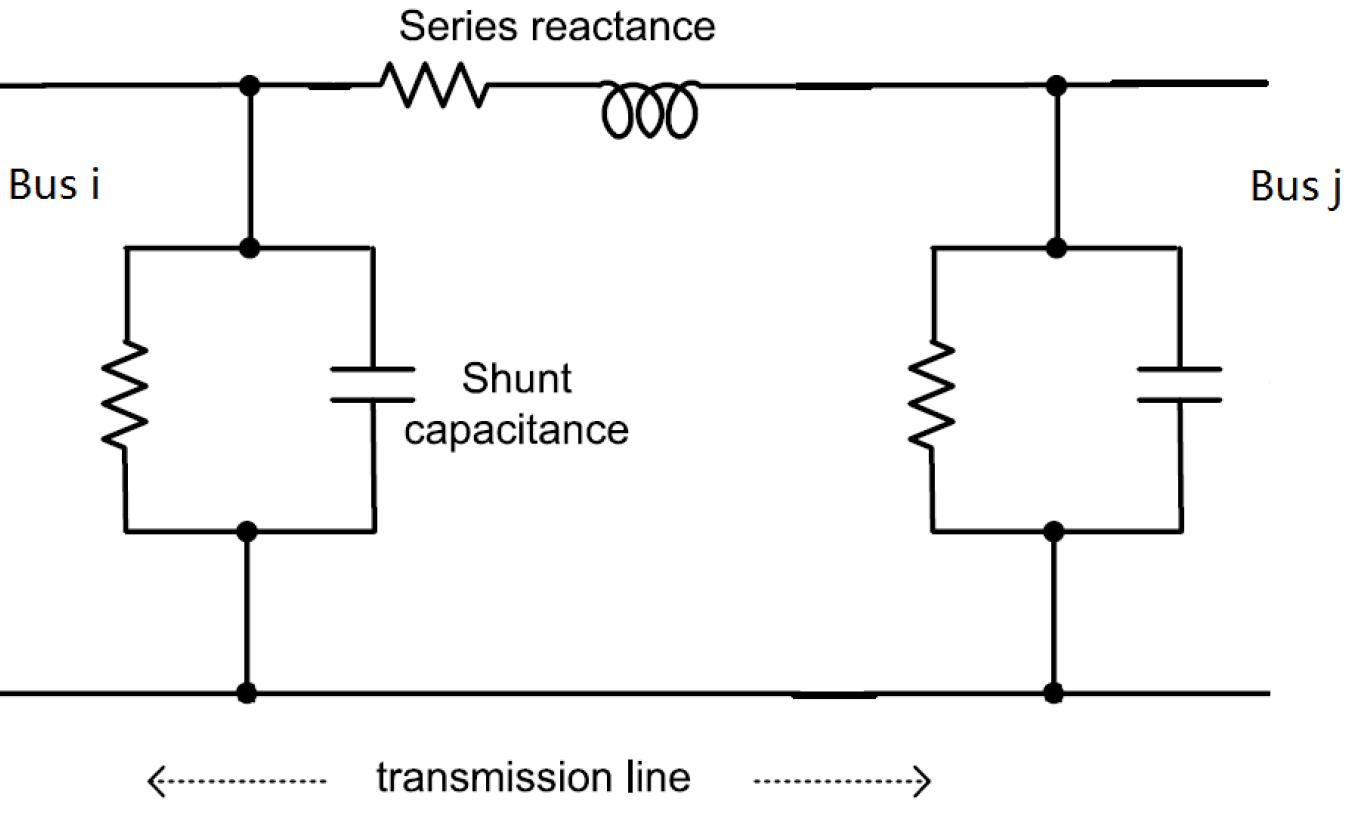}
\caption{Pi model of high voltage transmission line}
\label{pimodel}
\end{figure}

\indent A general electricity power systems consists of generators, loads and power distribution systems. Generations and loads are connected by power distribution system. Power system is accessed through high voltage transformers that raise their voltage to power transmission level. In power grid models, these transformers are represented by buses (or feeders) as a generation of interface between power transmission system and end generations or loads.
% As high voltages are crucial for power transmission over long distances to minimize power loss along the transmission line, power distribution system transmit electricity power at high voltage on the order of hundreds of kilovolts. Large commercial customers can directly receive electricity power at high voltage. And the generating sites are equipped with transformers that raises their voltage to that the magnitude that can directly access to the distribution system. 
Figure 2 shows the basic structure and components of a transmission and distribution system.
%The vertical lines represent buses, or common connections at key points in the system, especially power plants and large load access point. With power flowing from left to right, the diagram indicate the topology of the hierarchical relationship among the connection of the key points. End users in the subsystem connected to the buses by transformers which is a device for elevating and dropping the voltage in an AC circuit with a fixed magnification ratio. Different buses in the power system are connected by transmission lines.\\
%\begin{figure}[b]
%\centering
%\includegraphics[width=3.0in]{./Figures/power_distribution_network.eps}
%\caption{Basic structure and components of a transmission and distribution system.}
%\label{powergrid}
%\end{figure}
 In describing transmission-line parameters, pi model is widely adopted to consider inductance in series and capacitance in parallel. Figure (\ref{pimodel}) illustrates the modeling of a transmission line.\\
%Without delving into the details, we can use a pi model to characterize a line of an equivalent resistance, inductance and capacitance. The ratio of series impedance and shunt admittance determines a quantity called the characteristic impedance. Generally the overall impedance of a line tends to be dominated in practice by its inductive reactance.\\
\indent Based on the previous background, the power network can be described in so termed network matrix model \cite{du2013network, AsynNetwork}. Let $\mathcal{V}\triangleq\{0,1,2....N\}$ denotes the set of all buses, $\boldsymbol{u}\triangleq[u_{0},u_{1}...u_{N}]^{T}$ refers to the vector of voltage at each buses, $\boldsymbol{i}\triangleq[i_{0},i_{1}...i_{N}]^{T}$ refers to the vector of current at each buses. Each line is characterized by the admittance matrix $\boldsymbol{Y}=[Y_{n,m}]_{N*N}$, whichs include line admittance $Y_{n,m}=G_{n,m}+iB_{n,m}$ and shunt admittance admittances $\bar{Y}_{n,m}=\bar{G}_{n,m}+i\bar{B}_{n,m}$ in the pi-model of line ${n,m}\in\xi$, and self-admittance $Y_{n,n}=-\sum_{m\neq n}(\bar{Y}_{n,m}+Y_{n,m})$. As the shunt capacitance is so small relative to line impedance, we often neglect the shunt admittance part in network admittance matrix. $\boldsymbol{u}$ and $\boldsymbol{i}$ have the following relationship \cite{du2014distributed, du2014distributed}\\
\begin{equation} \label{admittant matrix}
\boldsymbol {i}=\boldsymbol{Y}\boldsymbol{u}.
\end{equation}
This relationship is developed by applying KCL at each bus in the system to transform Olm's equation regarding the bus voltage and line current's to bus voltage and bus current injection relation.

%\indent We model nominal apparent power $s_{v}$ the loads and renewable distributed generators via the following law relating voltage $u_{v}$ and current $i_{v}$.
\indent The apparent power $s_{v}$ on bus v has the following relation ship with its voltage and current injection.\\
\begin{equation}
s_{v}=u_{v}i^{*}_{v},
\end{equation}
where $s_{v}=p_{v}+iq_{v}$, $p_{v}$ is the active power and $q_{v}$ is the reactive power.
%Actually the positive active power $p_{v}$ is average net electrical power that is supplied to the load .
%Figure \ref{recpower} shows the physical meaning of active power and reactive power based on the time domain instantaneous power consumption. \\
%\indent Let $\boldsymbol{p}_{all}\triangleq[p_{0},p_{1}...p_{N}]^{T}$ denotes the vector of real power injection at each buses, $\boldsymbol{q}_{all}\triangleq[q_{0},q_{1}...q_{N}]^{T}$ denotes the vector of reactive power injection at each buses and $\boldsymbol{s}_{all}\triangleq[s_{0},s_{1}...s_{N}]^{T}$ denotes the vector of apparent power injection at each bus. \\
%\begin{figure}[b]
%\centering
%\includegraphics[width=3.0in]{./Figures/rective_power1.eps}
%\caption{Power as a product of voltage and current, with current lagging voltage by a phase of $\phi$. The power curve in time domain is shifting down from the one with the same current and voltage phase. $P_{ave}$ is the active power consumption averaged in time. $S$ is the apparent power, where reactive power is $\sqrt{S^2-P_{ave}^2}$.}
%\label{recpower}
%\end{figure}
\subsection{Linear Approximation for MicroGrid}
\indent We adopt the linear approximative model developed by S. Bolognani   as the system model for our MiroGrid. This model is derived by taking the first order Tyler approximation form of the bus voltage as a function of bus power injection. The model is briefly described as follows:\\
\indent We define matrix $\boldsymbol{X}$ as
\begin{equation}
\begin{bmatrix}
\boldsymbol{X} & \boldsymbol{1}         \\[0.3em]
\boldsymbol{1}^T   & 0            \\[0.3em]
\end{bmatrix}
=
\begin{bmatrix}
\boldsymbol{Y} & \boldsymbol{1}         \\[0.3em]
\boldsymbol{1}^T   & 0            \\[0.3em]
\end{bmatrix}
^{-1},
\end{equation}
where $\boldsymbol{Y}$ is the admittance matrix defined in previous subsection.\\
\indent The relationship between magnitude of voltage at the each bus $\boldsymbol{u}$ and the power injection at each bus $\boldsymbol{s}$ is given by\\
\begin{equation}
\label{bus voltage eq}
 |u_{v}| \approx U_{N}+\frac{1}{U_{N}}[\Re([\boldsymbol{X}]_{i\notin 0}\boldsymbol{s})]_v,
\end{equation}
and the total power loss on the transmission line are given by\\
\begin{equation}
\label{power loss1}
p_{\boldsymbol{l}}=\frac{1}{U_N^2}({\bf p}^{T}\Re([\boldsymbol{X}]_{i\notin 0,j\notin 0}){\bf p}+{\bf q}^{T}\Im([\boldsymbol{X}]_{i\notin 0,j\notin 0}){\bf q},
\end{equation}
 where $\boldsymbol{s}=\boldsymbol{p}+i\boldsymbol{q}\triangleq[s_{1}...s_{N}]^{T}$
%  The detailed description of this linear approximation model will be given in Appendix A.\\

%\subsection{Specific Model for Our MicroGrid}
%\indent Consider our MicroGrid consists of $n+1$ buses. These buses are numbered in set $\mathcal{V}=\{0,1,2....N\}$. Let the bus 0 be the PCC bus that is connected to grid. Bus 0 is a slack bus with fixed voltage magnitude and phase of  $U_{N}e^{j\phi}$. The renewable generating is distributed among the buses $v\in \mathcal{V}_{g}$. The distributed generating (DG) bus have a time-varying passive power harvesting $p_{g}(v,t)$. Smart buildings can be considered as linear resistive load. We assume that smart building at bus $v\in \mathcal{V}_{c}$ actively adjust its power consumption $p_{c}(v,t)$ without the influence of the operating condition of the MicroGrid, as the voltage keeps stabilized in a small range. Reactive power consumption is neutralized by the reactive compensation unit that is installed as a standard for smart building. For simplicity, we assume that the MicroGrid only consists of these two kinds of buses apart from bus 0. We also assume that bus $v\in \mathcal{V}_{c}$ only provide electrical energy for smart buildings.\\
\indent Consider our MicroGrid consists of $N+1$ buses. These buses are numbered in set $\mathcal{V}=\{0,1,2....N\}$. The renewable generating is distributed among the buses $v\in \mathcal{V}_{g}$ with a time-varying passive power harvesting $p_{g}(v,t)$. Smart buildings is located on bus $v\in \mathcal{V}_{c}$ with adjustable power consumption $p_{c}(v,t)$. Our MicroGrid connects to grid through point of common coupling (PCC). Let the bus 0 be the PCC bus with fixed voltage $U_{N}e^{j\phi}$ as the requirement of current prevailing interface standard. Reactive power consumption is neutralized by the reactive compensation unit. For simplicity, we assume that the MicroGrid only consists of these two kinds of buses apart from bus 0. We also assume that bus $v\in \mathcal{V}_{c}$ only provide electrical energy for smart buildings.\\
\indent Specifically in our model, The vector of power harvesting at power generation bus are denoted as:
\begin{equation} \label{model notation1}
\boldsymbol{p}_g(t)=[p_1(t),p_2(t),...p_{N_g}(t))]^T, \hspace{0.2em}         \mathcal{V}_{g}=\{1,2,..N_g\}.
\end{equation}
\indent The vector of power consumption at consumption buses that are connected to the smart building are denoted as:
\begin{equation} \label{model notation2}
\begin{gathered}
\boldsymbol{p}_c(t)=[-p_{{N_g}+1}(t),-p_{{N_g}+2}(t),...-p_{N}(t))]^T, \hspace{0.2em} \\
\mathcal{V}_{c}=\{{N_g}+1,{N_g}+2,..N\}.
\end{gathered}
\end{equation}
\indent So that the power vector $\boldsymbol{p}_c(t),\boldsymbol{p}_g(t),$ at each bus can be written as the component of $\boldsymbol{s}_{/}(t)$
\begin{equation}  \label{model notation3}
\boldsymbol{p}(t)=[\boldsymbol{p}_{g}(t),-\boldsymbol{p}_{c}(t)]^T.
\end{equation}
As reactive power is required to compensate at each buses by current MicroGrid standard, we only consider the active power at each buses.
\begin{equation}  \label{model notation3}
\boldsymbol{s}(t)=\boldsymbol{p}(t).
\end{equation}
The matrix X can be composed to the sub-matrix that corresponds to the dimension of $N_g$ and $N_c$
\begin{equation}
\Re(\boldsymbol{X})=\begin{bmatrix}
0 & 0 &    0    \\[0.3em]
0  & \boldsymbol{M}     &\boldsymbol{N} \\[0.3em]
0& \boldsymbol{N}^T & \boldsymbol{Q}      \\
\end{bmatrix},
\end{equation}
 with $M\in \mathbb{R}^{N_g \times N_g}$ , $N\in \mathbb{R}^{N_g \times N_c}$ and $Q\in \mathbb{R}^{N_c \times N_c}$.
\indent By substituting the decomposition of matrix X (\ref{Xdec}) and power consumption vector $\boldsymbol{p}(t)$ (\ref{model notation3}) into power loss equation (\ref{power loss1}), we have
\begin{equation}
\label{Xdec}
\begin{aligned}
p_{\boldsymbol{l}}(t)
%&=\frac{1}{U_N^2}({\boldsymbol{p}_{/}}^{T}\Re({\boldsymbol{[X]}_{i,j\notin0}}){\boldsymbol{p}_{/}}+{\boldsymbol{q}_{/}}^{T}\Im({\boldsymbol{[X]}_{i,j\notin0}}){\boldsymbol{q}_{/}})
%\\&=\frac{1}{U_N^2}{\boldsymbol{p}_{/}}^{T}\Re({\boldsymbol{[X]}_{i,j\notin0}}){\boldsymbol{p}_{/}}
&=[\boldsymbol{p}_{g}(t)^T,-\boldsymbol{p}_{c}(t)^T]
\begin{bmatrix}
 \boldsymbol{M}     &\boldsymbol{N} \\[0.3em]
 \boldsymbol{N}^T & \boldsymbol{Q}      \\
\end{bmatrix}
[\boldsymbol{p}_{g}(t)^T,-\boldsymbol{p}_{c}(t)^T]^T\\
&=\frac{1}{U_N^2}(\boldsymbol{p}_{g}(t)^T\boldsymbol{M}\boldsymbol{p}_{g}(t)-2\boldsymbol{p}_{g}(t)^T\boldsymbol{N}\boldsymbol{p}_{c}(t)+\boldsymbol{p}_{c}(t)^T\boldsymbol{Q}\boldsymbol{p}_{c}(t)).
\end{aligned}
\end{equation}
\indent Similarly by substituting the decomposition of matrix X (\ref{Xdec}) and power consumption vector $\boldsymbol{p}(t)$ (\ref{model notation3}) into bus voltage magnitude equation (\ref{bus voltage eq}), we have
\begin{equation}
\begin{aligned}
|\boldsymbol{u}|
%&= U_{N}+\frac{1}{U_{N}}\Re([\boldsymbol{X}]_{i\notin 0}\boldsymbol{s}_{/})\\
%&= U_{N}+\frac{1}{U_{N}}\Re[\boldsymbol{X}]_{i\notin 0}\boldsymbol{p}_{/}
%\\&= U_{N}+\frac{1}{U_{N}}\Re[\boldsymbol{X}]_{i\notin 0}\boldsymbol{p}_{/}
&= U_{N}+\frac{1}{U_{N}}
\begin{bmatrix}
  \boldsymbol{M}     &\boldsymbol{N} \\[0.3em]
\boldsymbol{N}^T & \boldsymbol{Q}
\end{bmatrix}
[\boldsymbol{p}_{g}^T,-\boldsymbol{p}_{c}^T]^T
\\&= U_{N}+\frac{1}{U_{N}}[(\boldsymbol{M}\boldsymbol{p}_{g}(t)-\boldsymbol{N}\boldsymbol{p}_{c}(t))^T,(\boldsymbol{N}^T\boldsymbol{p}_{g}(t)-\boldsymbol{Q}\boldsymbol{p}_{c}(t))^T]^T.
\end{aligned}
\end{equation}
%%%%%%%%%%%%%%%%%%%%%%%%%%
%
% Convergence Analysis
%we first develop the concept of the objective profit. We then go extensively to give a concrete notation of the objective profit as a function of power injection at each bus.\\
%%%%%%%%%%%%%%%%%%%%%%%%%%%%%%%%%%%%%%%%%%%%%%%%%%%%%%%%%%%
\section{Problem Formulation}
\label{problem formulation}
\indent In this section, we first introduce our framework of optimal power management in renewable driven smart building MicroGrid with renewable generation in a systematic scale. We then develop our user satisfaction and electricity consumption balanced (USECB) profit model with the constrain region of power network in a concrete notation of a function of power injection at each bus. Based on our USECB model, objective profit is quantified as a measurement of the overall achievement which is user satisfactory related. Our power management strategy is based on maximizing the net profit in our USECB model as an objective function subject to the constraint of power capacity and power network safe operation limit. Finally, we model the practical problem in noise contaminated environment as a stochastic optimization problem aiming at minimizing the expectation of a stochastic function.
\subsection{Smart Building Power Management Overview}
\indent As the majority of devices integrated in the MicroGrid are power electronic based, MicroGrid is smartened by electronic systems' smooth and accurate control ability and fast computation power. Thanks to the state-of-arts power electronic system, system cores namely supervisory control and data acquisition system (SCADA) and user end devices such as distributed sensors that are monitoring the dynamic status of the user's systems and power management modules that are smartly controlling and adjusting user's loads are highly integrated in the controlling system to give MicroGrid a high degree of flexibility to act as a single controlled unity. This makes it possible for us to apply a centralized management strategy rather than a distributed one as the low complexity of the small MicroGrid system, high robustness of the control system and the little transmission delay in the whole system. A centralized control unit exhibits as the core for gathering information about the overall system's status and generating control signals based on the assembled optimization programs.\\
\indent At the beginning of each time slot, information from distributed sensors are gathered at the central control unit. The information gathered at the central control unit are real time inside temperature of each building at bus v $c_{in}(v,t)$, outside temperature of each building at bus v $c_{out}(v,t)$, power injection $p_{g}(v,t)$ at each generating bus $v \in \mathcal{V}_{g}$. The centralized control unit make real time response by adjusting the load on consumption buses. This decision is based on maximizing an objective function of our USECB model without harming the system stability by fluctuating bus voltage too much. The controlling strategy utilize in time censoring data while the previous result is also considered.
\subsection{Quantifying Objective Profit}
\indent We put forward the concept of net profit in our USECB model that achieves the compromising between power reliance on outside grid and user's satisfaction maximizing. To incorporate the concept of our objective profit in USECB model in the framework of real-time power management, the net profits in our USECB model are evaluated on a per time slot bases.\\
\begin{mydef}
 Given the air conditioning related flexible power consumption at each bus $\boldsymbol{p}_c(t)$ (the concept of air conditioning related flexible power consumption will be explained as follows), the USECB net profit $\pi(\boldsymbol{p}_c(t))$ is defined as user's overall satisfaction related revenue minus its power cost from outside commercial grid.
\end{mydef}

\indent We explore the modeling and computation of USECB in the following part. We first develop the concept of AC related flexible power consumption. Though the total power consumption can be of multiple types such as AC systems, venting, lighting and absolute user demand. In this paper, we only consider adjusting the load that supplies AC systems and study the effect of managing these kinds load supplied to AC systems as other types of power consumption have limited degree of flexibility or its impact on the overall system is negligible.\\
\indent Given the power consumption $p_c(v,t)$ at time t for the smart building that are connected to bus v, we denote the satisfactory revenue of its users as $U(p_c(v,t),t)$ by cooling the indoor temperature closer to user's set level. We also denote $\lambda$ as the unit electricity price for power taken from outside commercial grid. Therefore the net profit in USECB model can be expressed in the following form:\\
\begin{equation}
\pi(\boldsymbol{p}_c(t),t)=\sum_{v \in \mathcal{V}_g}U(p_c(v,t),t)-\lambda p_0(t).
\end{equation}
\indent {\bf User Satisfaction Related Revenue}
\indent In many research papers such as \cite{chen2012optimal, li2011optimal}, the utility function of AC devices is usually set to be proportional to the square of difference between the actual inside temperature and set temperature.
\begin{equation} \label{general satisfy}
U(p_c(v,t),t)=-\beta(c_{in}(v,t+1)-c_{set}(v))^{2}.
\end{equation}
where $c_{set}(v)$ is the objective goal set by user in bus v. User in bus v wish the temperature to be adjusted around the set temperature $c_{set}(v)$  at which the user will feel most satisfied. $c_{in}(v,t+1)$ is the predictive inside temperature in the next time slot which is sum up by the current inside temperature and heating coming from the outside and the cooling power refrigerated by AC. According to the thermodynamic equation, we have
\begin{equation} \label {thermalequation}
\begin{gathered}
(c_{in}(p_c(v,t),t+1)-c_{in}(v,t))C_{room}=\\
\lambda_{room}(c_{out}(v,t)-c_{in}(v,t))\Delta t+\eta', p_{c}(v,t)\Delta t.
\end{gathered}
\end{equation}
 where $C_{room}$ is specific heat capacity of the smart building, $\lambda_{room}$ is the heat transfer coefficient between smart building and outside, $\eta'$ is the cooling efficiency of the AC. By taking (\ref{thermalequation}) into (\ref{general satisfy}), we have the following relationship between the realtime AC injection power and users satisfaction.
\begin{equation} \label{user statisfaction1}
\begin{aligned}
U(p_c(v,t),t)&=-\beta (c_{in}(v,t)+\frac{\lambda_{room}}{C_{room}}(c_{out}(v,t)-c_{in}(v,t))\Delta t\\
&+\frac{\eta' }{C_{room}}p_{c}(v,t)\Delta t-c_{set}(v))^{2}.
\end{aligned}
\end{equation}
\indent To simplify the model by eliminating the redundant variables, we set$\frac{\lambda_{room}}{C_{room}}=\alpha_{1}$ ,$\frac{\eta' p_{c}(v,t)}{C_{room}}=\alpha_{2}$ ,equation (\ref{user statisfaction1}) is written as follows:
\begin{equation} \label{users satisfaction2}
\begin{aligned}
U(p_c(v,t),t)&=-\beta (c_{in}(v,t)+\alpha_{1}(c_{in}(v,t)-c_{out}(v,t))\Delta t\\
&+\alpha_{2}p_{c}(v,t)\Delta t-c_{set}(v))^{2}.
\end{aligned}
\end{equation}
\indent The overall net revenue of all users written as the summation of user welfare in all buses in vector form $\boldsymbol{p}_c(t)$ can be denoted as:\\
\begin{equation}\label{objective function1}
\begin{aligned}
U(\boldsymbol{p}_c(t),t)&=\sum_{v\in \mathcal{V}_c} U(p_c(v,t),t)
\\&=-\beta (\boldsymbol{c}_{in}(t)+\Lambda_{1}(\boldsymbol{c}_{out}(t)-\boldsymbol{c}_{in}(t))\Delta t\\
&-\Lambda_{2}\boldsymbol{p}_{c}(t)\Delta t-\boldsymbol{c}_{set}(v))^{T}
(\boldsymbol{c}_{in}(t)
+\Lambda_{1}(\boldsymbol{c}_{out}(t)-\boldsymbol{c}_{in}(t))\Delta t\\
&-\Lambda_{2}\boldsymbol{p}_{c}(t)\Delta t-\boldsymbol{c}_{set}(t)),\\
\end{aligned}
\end{equation}
where $\Lambda_{1}=diag(\alpha_{1})$ and $\Lambda_{1}=diag(\alpha_{2})$\\
\indent {\bf Power Cost to Grid}
%Consider our mircogrid consists of $n+1$ buses. These buses are numbered in set $\mathcal{V}=\{0,1,2....N\}$. Let the bus 0 is the bus that connected to grid. Bus 0 is a slack bus. The voltage magnitude and phases in this bus is fixed at $U_{N}e^{j\phi}$. The renewable generating is distributed among the buses $v\in \mathcal{V}_{g}$. The distributed generating(DG) bus have a time-varying passive power harvesting $p_{g}(v,t)$. Smart buildings are connected to the buses $v\in \mathcal{V}_{g}$. Smart buildings can be considered as linear resistive load. We assume smart building at bus $v\in \mathcal{V}_{c}$ actively adjust its consuming power $p_{c}(v,t)$ without the influence of the operating condition of the microgrid, as the voltage keeps stabilized in a small range. Reactive power consumption is neutralized by the reactive compensation unit that is installed as a standard for smart building. For simplicity, we assume that the microgrid only consists of these two kinds of buses apart from bus 0. We also assume that bus $v\in \mathcal{V}_{c}$ only provide electrical energy for smart buildings.\\
\indent Apart from fixed cost for MicroGrid's basic maintenance, the major additional cost for MicroGrid is the cost to procure additional power from outside commercial grid. We assume the electricity cost is proportional to the power intake from the grid $p_0(t)$ at unit price of $\lambda$. Here we denote $p_0(t)$ as a function of $p_c(v,t)$ by the law of energy conservation and power loss equation 
\begin{equation}
\begin{aligned}
p_{0}(t)&=\boldsymbol{1}^T\boldsymbol{p}_{c}(t)-\boldsymbol{1}^T\boldsymbol{p}_{g}(t)\\
&+\frac{1}{U_N^2}(\boldsymbol{p}_{g}(t)^T\boldsymbol{M}\boldsymbol{p}_{g}(t)-2\boldsymbol{p}_{g}(t)^T\boldsymbol{N}\boldsymbol{p}_{c}(t)+\boldsymbol{p}_{c}(t)^T\boldsymbol{Q}\boldsymbol{p}_{c}(t)).\\
\end{aligned}
\end{equation}
\subsection{Load Constraint in MicroGrid}
%\indent The stable operation of the power grid system relies on little fluctuation of voltage in each buses. To maintain MicroGrid system stability, the power consumption at each buses should be restricted in the corresponding region so that the magnitude of voltage at each should be kept within the safe operation limit.
\indent To maintain MicroGrid system stability, the magnitude of voltage should be kept within safe operation limit.
\begin{equation} \label{voltage constraint}
v_{min}\leq |\boldsymbol{u}| \leq v_{max}.
\end{equation}
%Another constraint for power consumption at each bus is that it should not exceed the maximum power of AC devices,for each AC device has its physical maximal operation power constraint.We denote $p_{max}(v)$ as the maximum power consumption at bus v. \\
The power consumption at each bus should be remained within its capacity.
\begin{equation} \label{power constraint}
p_{min}\leq \boldsymbol{p}_{c}(t) \leq p_{max}.
\end{equation}
\indent We denote set $\mathcal{A}$ as the convex set of power constraint which satisfies the voltage stable condition and maximum power constrain.
\begin{equation} \label{constrain reigon}
\begin{aligned}
\boldsymbol{p}_{c}(t) \in \mathcal{A}:\{\boldsymbol{p}_{c}(t)\hspace{0.5em} |\hspace{0.5em} &v_{min}\leq |\boldsymbol{u}(U_{N}
,\boldsymbol{p}_{c}(t))| \leq v_{max}\hspace{0.5em}\\
&\cap p_{min}\leq \hspace{0.5em} \boldsymbol{p}_{c}(t)\leq p_{max} \hspace{0.5em} \}.
\end{aligned}
\end{equation}
\subsection{Power Management Problem in Noise Environment}
\indent As $\pi(\boldsymbol{p}_c(t),t)$ can be written as:\\
\begin{equation}
\pi(\boldsymbol{p}_c(t),t)=\lambda f(\boldsymbol{p}_c(t),t)+c,
\end{equation}
 where
\begin{equation}
\begin{aligned}
f(\boldsymbol{p}_c(t),t)&=\frac{\beta}{\lambda} [(\boldsymbol{p}_{c}(t)^T\Lambda_1\Lambda_2\boldsymbol{p}_{c}(t)\Delta t^2\\
&-2(\boldsymbol{c}_{in}(t)+\Lambda_1(\boldsymbol{c}_{out}(t)-\boldsymbol{c}_{in}(t))\Delta t-\boldsymbol{c}_{set}(t))^T\Lambda_2\boldsymbol{p}_{c}(t))\Delta t]\\
&+\boldsymbol{1}^T\boldsymbol{p}_{c}(t)+\lambda\frac{1}{U_N^2}(-2\boldsymbol{p}_{g}^T\boldsymbol{N}\boldsymbol{p}_{c}+\boldsymbol{p}_{c}^T\boldsymbol{Q}\boldsymbol{p}_{c})\\
c&=(\boldsymbol{c}_{in}(t)+\Lambda_1(\boldsymbol{c}_{out}(t)-\boldsymbol{c}_{in}(t))^T(\boldsymbol{c}_{in}(t)+\Lambda_1(\boldsymbol{c}_{out}(t)-\boldsymbol{c}_{in}(t))+\boldsymbol{1}^T\boldsymbol{p}_g.
\end{aligned}
\end{equation}
\indent Maximizing $\pi(\boldsymbol{p}_c(t),t)$ is equally as minimizing $f(\boldsymbol{p}_c(t),t)$
%\begin{equation} \label{problem notation}
%\begin{gathered}
%\min:\beta (\boldsymbol{p}_{c}(t)^Tdiag(\boldsymbol{\alpha}_{2})diag(\boldsymbol{\alpha}_{2})\boldsymbol{p}_{c}(t)\Delta t^2\\
%-2(\boldsymbol{c}_{in}(t)+diag(\boldsymbol{\alpha}_{1})(\boldsymbol{c}_{out}(t)-\boldsymbol{c}_{in}(t))\Delta t-\boldsymbol{c}_{set}(t))^Tdiag(\boldsymbol{\alpha}_{2})\boldsymbol{p}_{c}(t))\Delta t\\
%+\lambda\boldsymbol{1}^T\boldsymbol{p}_{c}(t)+\lambda\frac{1}{U_N^2}(-2\boldsymbol{p}_{g}^TN\boldsymbol{p}_{c}+\boldsymbol{p}_{c}^TQ\boldsymbol{p}_{c})\\
%s.t.: \boldsymbol{p}_{c}(t) \in \mathcal{A}:\\
%p_{min} \leq \boldsymbol{p}_c(t) \leq p_{max} \hspace*{1em} ,\forall v\in \mathcal{V}_{c}\\
%v_{min} \leq U_{N}+\frac{1}{U_{N}}[(M\boldsymbol{p}_{g}-N\boldsymbol{p}_{c})^T,(N^T\boldsymbol{p}_{g}-Q\boldsymbol{p}_{c})^T]^T \leq v_{max}
%\\
%\end{gathered}
%\end{equation}
%\indent Optimizing the power consumption $\boldsymbol{p}_{c}(t)$ with the exact value of other parameters $f(\boldsymbol{p}_c(t),t)$ is a standard convex optimization problem. Many well developed optimization algorithm such as ellipsoid method and interior point method can be applied directly into this kind of problems. However as these parameters are obtained in real time measurement, the exact data is always contaminated by noise. The challenge in our problem formation is that we are optimizing the power consumption $\boldsymbol{p}_{c}(t)$ without knowing the exact value of the parameters in the objective function.\\
\indent As the data obtained from the distributed sensors are always the noise corrupted version of the their actual value, we consider our the optimal smart building power management problem in noise contaminated environment as a stochastic optimization problem in statistical point of view. We observe function $f(\boldsymbol{p}_{c}(t);\hat{Z})$ in multiple perspective. $f(\boldsymbol{p}_{c}(t);\hat{Z})$ is notated as a clear deterministic analytical form as a function of $\boldsymbol{p}_{c}(t)$ and $Z$, where $Z$ is the parameter of $f$ namely $\boldsymbol{c}_in$, $\boldsymbol{c}_out$ and $\boldsymbol{p}_g$. However these parameters in Z appears in the form of random variables $\hat{Z}\sim P$ when noise is introduced. In this way, $f(\boldsymbol{p}_{c}(t);\hat{Z})$ will be a random variable of which distribution is a function of $\boldsymbol{p}_{c}(t)$. $f(\boldsymbol{p}_{c}(t);\hat{Z})$ is no longer a deterministic function of $\boldsymbol{p}_{c}(t)$. Instead random variables $f(\boldsymbol{p}_{c}(t);\hat{Z})$ only follows the distribution that is determined as a function of $\boldsymbol{p}_{c}(t)$. However its statistical properties such as expectation is still a deterministic function of $\boldsymbol{p}_{c}(t)$. Under the previous context, we model the problem of optimal smart building power management as minimizing the expectation of a randomized function rather than its exact value.
%\indent In statistical point of view, we denote the measurement value at one time slot as a sample of random variable $Z\sim P$. The objective function parameterized by the random sample is denoted as \\
\begin{equation} \label{problem notation}
\begin{gathered}
\min:\mathbb{E}[f(;\hat{Z})]\\
s.t : \boldsymbol{p}_{c}(t) \in \mathcal{A}.
\end{gathered}
\end{equation}
\section{Algorithm Design and Convergence Analysis}
\label{stochastic solver}
\indent As noise corrupted environment impedes us from deriving any exact analytical form of the expectation function, this rules out us to find any analytical solution or applying approximative approaches in convex optimization directly. Therefore theoretical upper bound is not achievable when the measurement is corrupted by noise. Indeed we exploit Bregmen projection based mirror decent algorithm to iterate the objective variable $\boldsymbol{p}_{c}(t)$ that gradually converges to the optimization point of the expectation functions. In this section, we first introduce our Bregmen projection based mirror decent algorithm  and applies our algorithm to the specific problem of optimal power management in smart building in noise corrupted environment. Convergence analysis is provided in the final part of this section as a theoretical base.  The notations and discussions in convergence analysis part are made intentionally independent of other parts of the paper in order to present the proof in a mathematically general way \cite{duchi2010composite, bubeckintroduction}.
\subsection{Stochastic Approximation Solver}
\indent We begin introducing our Bregmen projection based mirror decent algorithm by first presenting the definition of Bregmen function and divergence. The Bregmen divergence associated with $\psi$ is defined as:\\
\begin{mydef} Let $\psi$ denotes a continuous differentiable function. The Bregmen divergence associated with $\psi$ is defined as:\\
\begin{equation}
B_{\psi}(\boldsymbol{\omega,v})=\psi(\boldsymbol {\omega})-\psi({\bf v})-\langle \bigtriangledown\psi({\bf v}),\boldsymbol{\omega-v}\rangle.
\end{equation}
\end{mydef}
%
%\indent We then define the strongly convex property of a Bregemen function.\\
%\begin{mydef} Let $\psi$ denotes a continuous differentiable function that is $\alpha$-strongly convex, $\Vert \cdot \Vert$ denotes the standard Eculid norm Bregmen. The Bregmen divergence $B_{\psi}$ associated with $psi$ has the following properties.\\
%\begin{equation}
%B_{\psi}(\boldsymbol{\omega,v})\geq \frac{\alpha}{2}\Vert \boldsymbol {\omega-v} \Vert^{2}
%\end{equation}
%\end{mydef}
%
%\indent Bregmen projection based mirror decent algorithm is mainly performed as two step iterations. The iteration step only access the full analytical form of one measurement of random function ${f(:,\hat{Z},)}$, the result is iterated along the sub-gradient of this measurement without considering the corresponding constrains.  The intermediate result is projected on the constrain region in the second step to complete the whole iteration process at time slot $t+1$. The online iterative result will gradually converge to the optimization point of the expectation functions in a few steps.\\
\indent Our online mirror decent method is described as the follows:\\
{\bf Step 1}: Online Mirror Decent
\begin{equation}
\label{iteration1}
\omega_{p^{0}_{c}}(t+1) =\bigtriangledown\psi^{d}(\bigtriangledown\psi(\boldsymbol{p}_{c}(t))-\eta \bigtriangledown f(\boldsymbol{p}_{c}(t))),
\end{equation}
where $\psi^{d}$ denotes the dual function of $\psi$.\\
\indent Specifically, we use square of Eculid norm $\psi(\boldsymbol{\omega})=\frac{1}{2}\Vert\boldsymbol{\omega}\Vert_{2}^{2}$ in our specific problem. The mirror decent step can written as follows in analytical form \cite{clockconf}:
\begin{equation}
\begin{aligned}
\label{iteration detail}
\omega_{p^{0}_{c}}(t+1) &=\boldsymbol{p}_{c}(t)-\eta (\Lambda_2\beta\Delta t(\boldsymbol{c}_{in}(t+1)+\Lambda_1(\boldsymbol{c}_{in}(t+1)-\boldsymbol{c}_{out}(t+1))\Delta t\\
&-\Lambda_2\boldsymbol{p}_{c}(t+1)\Delta t-\boldsymbol{c}_{set}(t+1))+\lambda\boldsymbol{1}+\lambda\frac{1}{U_N^2}(-2\boldsymbol{N}\boldsymbol{p}_{g}(t+1)+2\boldsymbol{Q}\boldsymbol{p}_{c})(t)).
\end{aligned}
\end{equation}
{\bf Step 2}: Bregemen Projection
\begin{equation}
\label{iteration2}
\boldsymbol{p}_{c}(t+1) =\arg \min_{x\in \mathcal{A}} B_{\psi}(x,\omega_{{p}^{0}_{c}}(t+1)).
\end{equation}
Comparing with the theoretical optimal solution $\boldsymbol{p}_{c}^{*}(t)$ at time slot t, the iterative result $\boldsymbol{p}_{c}(t)$ obtained by our algorithm has a utility loss of $\mathbb{E}(f(\boldsymbol{p}_{c}(t)))-\mathbb{E}(f(\boldsymbol{p}_{c}^*(t)))$. Under the assumption that the distribution of $f$ stays the same over time T, the accumulative utility loss over time T $R_{n}=\sum_{t=1}^{T}f(\boldsymbol{a_{t}})-\inf\sum_{t=1}^{T}f(\boldsymbol{a^{*}})$ can be upper bounded by a function of $O(\sqrt{T})$ which will be proved in the later part of this section at convergence analysis.

\subsection{Convergence Analysis}
\indent We recall that accumulative regret rate is defined as:
\begin{equation}
R_{n}=\sum_{t=1}^{T}f(\boldsymbol{a_{t}})-\inf\sum_{t=1}^{T}f(\boldsymbol{a^{*}}).
\end{equation}

%\indent ,where $Z_{t}$ is a filtration of random variable Z at time t, according to distribution P. The function $f_{t}(\cdot)=f_{t}(\cdot;Z_{t})$ that is accessible at time t is parameterized by $Z_{t}$. The objective optimization function  $f(\cdot)=\mathbb{E}(f(\cdot))\approx\frac{1}{T}\sum_{t=1}^{T}f_{t}(\cdot)$ is the expectation of the filtration $f_{t}(\cdot)$. According to the law of large number, in a sufficient large time span the expectation is equal to the mean of $f_{t}(\cdot)$ on time axis. $\boldsymbol{a}$ is the objective optimization variable of function $f(\cdot)$. In time slot t, the algorithm approaches $\boldsymbol{a}$ by iterating $\boldsymbol{a}_t$ in the step described by equation (\ref{iteration1}) and (\ref{iteration2}). $R_{n}$ indicates whether $\boldsymbol{a}_t$ converges to $\boldsymbol{a}$ and the convergence rate. We prove the convergence of the algorithm by giving an upper-bound of $R_{n}$.
 We prove the convergence of the algorithm by giving an upper-bound of $R_{n}$. Before going to the formal proof, we first introduce the three lemmas that our proof relies on.\\
%\indent Lemma 1 gives a similar property to the geometry of squared Eculidean distance that the Bregmen divergence have:\\
\begin{mylemma}
\indent  Let $\mathcal{A} \in \overline{\mathcal{D}}$ be a closed and convex set such that $\mathcal{A} \cap\mathcal{D}\neq\varnothing$. Then, $\forall x\in \mathcal{D}$,
\begin{equation}
b=\arg\min_{a\in{\mathcal{A}}}B_{\psi}(a,x),
\end{equation}
exists and is unique. Moreover $b\in \mathcal{A} \cap\mathcal{D}$ ,and $\forall a \in \mathcal{A}$,
\begin{equation}
B_{\psi}(a,x)\geq B_{\psi}(a,b)+B_{\psi}(b,x).
\end{equation}
\end{mylemma}
\begin{mylemma}
\begin{equation}
B_{\psi}(x,y)+B_{\psi}(y,z)=B_{\psi}(x,z)+\langle x-y,\psi'(z)-\psi'(y)\rangle.
\end{equation}
\end{mylemma}
\begin{mylemma}
\begin{equation}
B_{\psi}(x,y)-B_{\psi}(z,y)=\psi(x)-\psi(z)+\langle x-z,\psi'(x)-\psi'(z)\rangle.
\end{equation}
\end{mylemma}
%\indent Lemma 2 and Lemma 3 can be easily proved by the definition of Bregemen divergence.\\
%\indent The following result gives an upper bound of the probability based convergence rate of our algorithm:\\
\begin{mytheorem}
\indent Assuming that $\psi$ is $\alpha$ convex.The dual norm of subgradient satisfies $\Vert f'({\boldsymbol{\omega}})\Vert_{*}\leq G_{*}$ ,for any $\boldsymbol{\omega}$.And the iterative point $\boldsymbol{a_{t}}$ is set as close to the optimization point as for any $B_{\psi}(\boldsymbol{a_{*}},\boldsymbol{a_{t}})<D^{2}$.And the iteration step is set as $\eta_{t}=\frac{D\sqrt{\alpha}}{G_{*}\sqrt{t}}$.Then we have the following convergence bound.
\begin{equation}
P(R_{n}\geq \frac{2DG_{*}\sqrt{T}}{\sqrt{\alpha}}+\varepsilon)\leq \exp(-\dfrac{\alpha\varepsilon^{2}}{16TD^{2}G_{*}^{2}}).
\end{equation}
\end{mytheorem}
\begin{myproof}
By the definition of subgradient,we have
\begin{equation}
\begin{aligned}
&f(\boldsymbol{a_{t}})-f(\boldsymbol{a^{*}})\\
\leq &\langle \boldsymbol{a_{t}}-\boldsymbol{a^{*}},f'_{t}(\boldsymbol{a_{t}})\rangle+\langle\boldsymbol{a_{t}}-\boldsymbol{a^{*}},f'(\boldsymbol{a_{t}})-f'_{t}(\boldsymbol{a_{t}})\rangle.\\
\end{aligned}
\end{equation}
 Using the iteration equation (\ref{iteration1}), we have
\begin{equation}
\begin{aligned}
&\langle \boldsymbol{a_{t}}-\boldsymbol{a^{*}},f'_{t}(\boldsymbol{a_{t}})\rangle\\
= & \frac{1}{\eta_{t}}\langle\boldsymbol{a}-\boldsymbol{a}_t,\psi'(\boldsymbol{w}_{t+1})-\psi'(\boldsymbol{a}_{t})\rangle.\\
\end{aligned}
\end{equation}
\indent Applying Lemma 1 and Lemma 2, we have\\
\begin{equation}
\begin{aligned}
& \frac{1}{\eta_{t}}\langle\boldsymbol{a}-\boldsymbol{a}_t,\psi'(\boldsymbol{w}_{t+1})-\psi'(\boldsymbol{a}_{t})\rangle\\
= &\frac{1}{\eta_{t}}\langle a-a_t,\psi'(w_{t+1})-\psi'(a_{t})\rangle\\
= &\frac{1}{\eta_{t}}(B_{\psi}(a,a_{t})+B_{\psi}(a_{t},w_{t+1})-B_{\psi}(a,w_{t+1}))\\
\leq &\frac{1}{\eta_{t}}(B_{\psi}(a,a_{t})+B_{\psi}(a_{t},w_{t+1})-B_{\psi}(a,a_{t+1})-B_{\psi}(a_{t+1},w_{t+1})).\\
\end{aligned}
\end{equation}
\indent To sum up, we have
\begin{equation}
\begin{aligned}
&\sum_{t=1}^{T}\langle\boldsymbol{a_{t}}-\boldsymbol{a^{*}},f'_{t}(\boldsymbol{a_{t}})\rangle\\
\leq &\sum_{t=1}^{T} \frac{1}{\eta_{t}}(B_{\psi}(a,a_{t})+B_{\psi}(a_{t},w_{t+1})-B_{\psi}(a,a_{t+1})-B_{\psi}(a_{t+1},w_{t+1}))\\
=  &\frac{1}{\eta_{1}}B_{\psi}(a,a_{1})+\sum_{t=2}^{T}(\frac{1}{\eta_{t}}-\frac{1}{\eta_{t-1}})B_{\psi}(a,a_{t})+\sum_{t=1}^{T}\frac{1}{\eta_{t}}(B_{\psi}(a_{t},w_{t+1})-B_{\psi}(a_{t+1},w_{t+1})).
\end{aligned}
\end{equation}
\indent Applying Lemma 3,we have\\
\begin{equation}
\begin{aligned}
&B_{\psi}(a_{t},w_{t+1})-B_{\psi}(a_{t+1},w_{t+1})\\
=&\psi(a_{t})-\psi(a_{t+1})+\langle\boldsymbol{a_{t+1}}-\boldsymbol{a_{t}},\psi'(\boldsymbol{\omega_{t+1}})\rangle.
\end{aligned}
\end{equation}
\indent Applying the $\alpha$ convex property of Bregmen Divergence
\begin{equation}
\begin{aligned}
&\psi(a_{t})-\psi(a_{t+1})+\langle\boldsymbol{a_{t+1}}-\boldsymbol{a_{t}},\psi'(\boldsymbol{\omega_{t+1}})\rangle\\
\leq &\langle\boldsymbol{a_{t}}-\boldsymbol{a_{t+1}},\psi'(\boldsymbol{\omega_{t}})\rangle-\frac{\alpha}{2}\Vert a_{t}-a_{t+1}\Vert^{2}+\langle\boldsymbol{a_{t}}-\boldsymbol{a_{t+1}},\psi'(\boldsymbol{\omega_{t+1}})\rangle\\
=&-\eta_{t}\langle f'(a_{t}),a_{t}-a_{t+1}\rangle-\frac{\alpha}{2}\Vert a_{t}-a_{t+1}\Vert^{2}\\
\leq &\eta_{t} G_{*}\Vert a_{t}-a_{t+1}\Vert-\frac{\alpha}{2}\Vert a_{t}-a_{t+1}\Vert^{2}\\
\leq & \frac{(\eta_{t}G_{*})^{2}}{2\alpha}.
\end{aligned}
\end{equation}
\indent Let $\mathcal{F}_{t}$ be a filtration of with $Z_{\tau} \in \mathcal{F}_{t}$ for $\tau \leq t$.Since $\boldsymbol{w}_{t} \in \mathcal{F}_{t-1}$
\begin{equation}
\mathbb{E}[\langle\boldsymbol{a_{t}}-\boldsymbol{a^{*}},f'(\boldsymbol{a_{t}})-f'_{t}(\boldsymbol{a_{t}})\rangle|\mathcal{F}_{t-1}]=\langle\boldsymbol{a_{t}}-\boldsymbol{a^{*}},f'(\boldsymbol{a_{t}})-\mathbb{E}[f'_{t}(\boldsymbol{a_{t}})|\mathcal{F}_{t-1}]\rangle=0.
\end{equation}
\indent Thus $\sum_{t=1}^{T}\langle\boldsymbol{a_{t}}-\boldsymbol{a^{*}},f'(\boldsymbol{a_{t}})-f'_{t}(\boldsymbol{a_{t}})\rangle$ is a martingale difference sequence
\begin{equation}
\begin{aligned}
&\langle\boldsymbol{a_{t}}-\boldsymbol{a^{*}},f'(\boldsymbol{a_{t}})-f'_{t}(\boldsymbol{a_{t}})\rangle
\leq & \Vert \boldsymbol{a_{t}}-\boldsymbol{a^{*}}\Vert\Vert f'(\boldsymbol{a_{t}})-f'_{t}(\boldsymbol{a_{t}})\Vert_*
\leq & 2\sqrt{2/\alpha}DG_*.
\end{aligned}
\end{equation}
\indent Defining $\sum_{t=1}^{T}\langle\boldsymbol{a_{t}}-\boldsymbol{a^{*}},f'(\boldsymbol{a_{t}})-f'_{t}(\boldsymbol{a_{t}})\rangle=\gamma_{T}$.\\
\indent Applying Azuma's inequality, we have
\begin{equation} \label{azuma}
\begin{aligned}
&P(\gamma_{T}\geq \epsilon)
\leq &\exp(-\dfrac{\alpha\varepsilon^{2}}{16TD^{2}G_{*}^{2}}).
\end{aligned}
\end{equation}
\indent So that applying the previous derived inequality, finally we have
\begin{equation}
\begin{aligned}
&R_{n}\leq\sum_{t=1}^{T}(\langle\boldsymbol{a_{t}}-\boldsymbol{a^{*}},f'_{t}(\boldsymbol{a_{t}})\rangle +\langle\boldsymbol{a_{t}}-\boldsymbol{a^{*}},f'(\boldsymbol{a_{t}})-f'_{t}(\boldsymbol{a_{t}})\rangle)\\
\leq &\frac{1}{\eta_{1}}B_{\psi}(a,a_{1})+\sum_{t=2}^{T}(\frac{1}{\eta_{t}}-\frac{1}{\eta_{t-1}})B_{\psi}(a,a_{t})\\
&+\sum_{t=1}^{T}\frac{1}{\eta_{t}}(B_{\psi}(a_{t},w_{t+1})-B_{\psi}(a_{t+1},w_{t+1}))+\langle\boldsymbol{a_{t}}-\boldsymbol{a^{*}},f'(\boldsymbol{a_{t}})-f'_{t}(\boldsymbol{a_{t}})\rangle)\\
\leq  &\frac{D^{2}}{\eta_{T}}+\frac{G_{*}^{2}}{2\alpha}\sum_{t=1}^{T}\eta_{t}+\gamma_{T}\\
< &\frac{DG_{*}\sqrt{T}}{\sqrt{\alpha}}+\sum_{t=1}^{T}\frac{DG_{*}}{\sqrt{\alpha}}(\sqrt{t}-\sqrt{t-1})+\gamma_{T}\\
= &\frac{2DG_{*}\sqrt{T}}{\sqrt{\alpha}}+\gamma_{T}.
\end{aligned}
\end{equation}
\indent Applying inequality (\ref{azuma}), we finally get,
\begin{equation}
P(R_{n}\geq 2\frac{DG_{*}\sqrt{T}}{\sqrt{\alpha}}+\varepsilon)\leq \exp(-\dfrac{\alpha\varepsilon^{2}}{16TD^{2}G_{*}^{2}}).
\end{equation}
\end{myproof}

\section{Simulation}
\label{simulation}
\begin{figure}[b]
\centering
\includegraphics[width=3.5in]{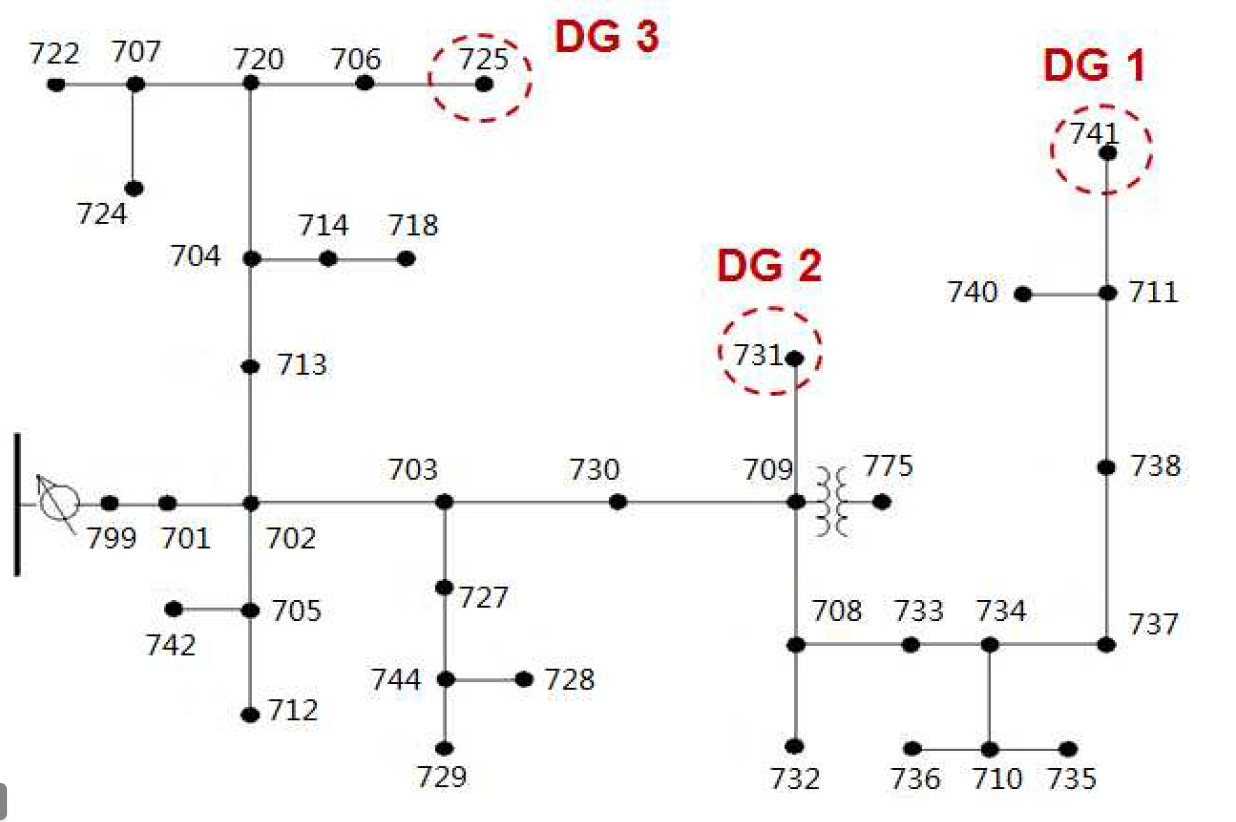}
\caption{The structure of IEEE 37 buses used in our simulation where distributed generations are on bus 725,731 and 741.}
\label{fig_exp1}
\end{figure}

\indent We test our novel stochastic power management approach on IEEE 37 buses which depicts the real power distribution network model from Southern California Edison. The topology of IEEE 37 buses is shown in Fig.(\ref{fig_exp1}).
% At the begining of each time slot, the regarding is collected to the central control. Always at the same time, the centralized control unit makes the optimal responce to manage power consumption on the consumption branches based on the datas collected from placed sensors. All the data collected at the sensors are the actual values of the data plus a white guassian noise. The deviation of guassian noise of each data is set as the $30\%$ of their actual value in our simulation. Here all the transmission and computation delays are omitted in our simulation as a result of the small scales of the power network systems.
 The ideal optimized power injection scheme is determined by solving the determinstic convex optimization problem (\ref{problem notation}) directly based on the exact data which provides a theoratical lower bound for the objective function (\ref{problem notation}). Performance is tested both by the result of actual power loss on the transmission line and power intake from out side grid. Two experiments are conducted in our whole simulation process.
 
 The dataset includes solar
 generation data gathered at three independent solar generation
 unit installed for the same home. In the dataset, the sensors
 monitors the amount of real time power generation at each
 unit in Augest 12th 2011 from 6am to 6pm which covers
 the full daytime generation range for PV. The normalized
 generation data is shown in Fig.(6). Our intension to choose
 the generation dataset from Smart Project  \cite{barker2012smart, chen2013non} are based on the
 following aspects. The dataset contains general patterns for
 PV unit daytime generation variation in a typical sunny day.
 Variations caused by sun rise and fall and fluctuations caused
 by clouds are obvious in this dataset. The dataset also includes
 different samples which reflects the variation in solar power
 harvesting at different sites that have the approximately same
 sunlight conditions.

In our simulation, we assume that three PV generation
sites of 12MW are placed respectively on bus 725,731 and
741. The normalized generation data from these three solar
generation units in Smart project are utilized in our simulation
to represent real-time PV generation variation on a per 48
sec bases. The root bus 799 is set as the PCC that is
connected to the grid. The remaining buses are set as the
consumption buses of which the power requirement includes
an infexible 0.6MW loads plus an adjustable loads that are
mainly driven by smart building cooling needs. The maxium
rate of central AC is set as 1.2MW at each buses under
which it can cool down the building at degrees per Tg per
minute. We neglect the power consumption of central AC
when it works in standby condition. Tg is set differently
following the distribution of N(1=pmax; 0:16=pmax) at each
consumption buses. Other thermostats parameters regarding
the smart buildings are obtained from real time indoor and
outdoor temperature data obtain from one smart home in Smart
project [18]. Renewable generation data is observed with a
white Gaussian Noise 30% of the actual value. Temperature
data is observed with a white Gaussian noise of variance 3.

In our first group of experiment, we are testing the convergence
property of our algorithm in ideal steady cases. The
static data is used in this experiment. Indoor temperature
is set fixed following the distribution of N(65; 5) in smart
buildings connected to each bus. Outdoor temperature and
renewable generation are set fixed at the data from Smart
project [18] at 7am. Stochastic scheme and exact scheme are
both tested in this simulation. Figure(7) shows the performance
of stchostic and exact method measured by the value of
objective minimization function. The red line is the exact
scheme comparing with our stochastic one in blue. The result
of our stochostic scheme converges in a few iteration steps
with a much better performance than the exact one which
highly fluctuates above our stochstic performance curve.

\begin{figure}
	\centering
	\includegraphics[width=4in]{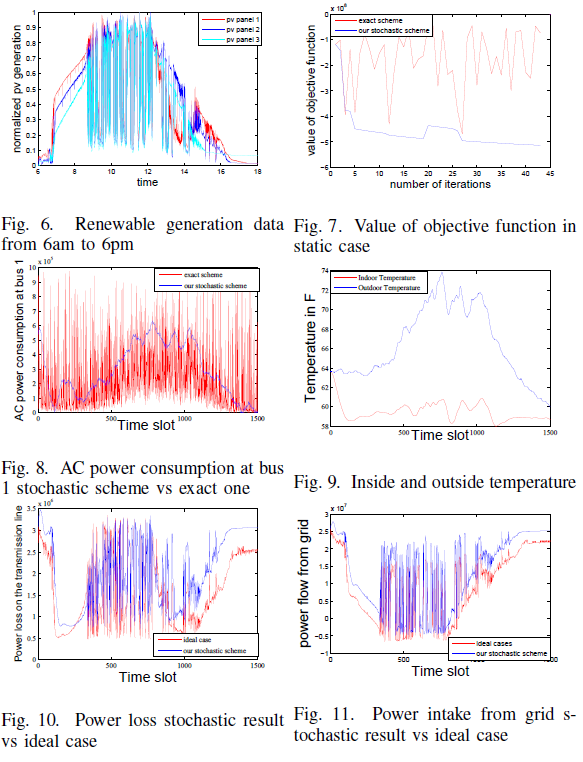}
	%\caption{The structure of IEEE 37 buses used in our simulation where distributed generations are on bus 725,731 and 741.}
\end{figure}

In our second group of experiment, we are testing our
algorithm in actual dynamic enviornment. The data collected
at each time slot is their real time actual value plus a white
Gaussian noise as described in previous context. Figure (8)
shows variation of air-conditioning power consumption at bus
701, where the blue line indicates our algorithm and the red
line indicates the exact scheme. Our algorithm has a much smoother
curve while the exact scheme adopts the noisy pattern
of the observed data. Figure (9) shows the actual temperature
variation of the smart building connected at bus 701. Temperatrue
is conditioned around the set goal in spite of the presence
of noise. Figure (10) shows the actual power loss on the
transmission line . The total power loss is reduced in the period
of high renewable penetration as the increased dependency on
loacl renewable which follows our assumption that effective
local power usuage reduces the load on the transmission line.
Figure (11) shows the actual power intake from the grid.
As the high renewable generation occures simulataneously
with high smart building power demand, load renewables are
nearly suffcient to supply smart building in peak hour. By
incorporating local renewables in MicroGrid, not only grid
capacity required for smart building is reduced but also the
comsumption pattern is inversed that smart building take less
powers in traditional peak demand period. Our algorithm can
approach the performance of the ideal schemes in most of
time in both of these two metrics. Overall the experiment has
tested the feasibility of our motivation of supplying the smart
building with local renewable, the effectiveness of our USECB
mode together with the effectiveness of our algorithm to filter
the noise, converge to the trend of actual data dynamics and
close to the performance in ideal case.

\section{Conclusions}\label{conclusion}
In this paper, we have outlooked the prospect of utilizing the local MicroGrid to drive the smart building and further investigated a optimal power management strategy. A USECB model has been presented to achieve balance between user welfare and external power consumption. Based the basis of proposed model, a Bregemen projection based mirror decent algorithm has been developed to optimize objective function in noise contaminated environment. Our algorithm has achieved an accumulative $O(\sqrt{T})$ loss than theoretical bound in steady state.\\
 %This model exhibits two main features of fairness. Firstly, it takes into consideration both practical needs of the maintenance of indoor temperature and efficient utilization of locally generated renewable energy. Secondly, the objective function denotes in a quadratic form which rarely adopts sparse solutions. This will result in relatively fair and even distribution of power injected to each buses.\\
%\indent On the basis of proposed model, we have been able to develop a stochastic, real time and iterative algorithm by leveraging the technique in the field of online machine learning. Our stochastic approximative slover only requires sensor information collected at current time slot and iterative result of last step as input. Superior to traditional deterministic convex optimization solver, our stochastic approximative solver only suffers an accumulative $O(\sqrt{T})$ loss comparing with optimal result in steady state under noise contaminated environment, while the traditional deterministic convex optimization solver can only achieve $O(T)$.\\
\indent We have tested the performance our algorithm in our simulation. We have proposed three metrics in our simulation which are power losses, external power consumption and objective function values. In the period of high local renewable penetration, power losses and external power consumption have been both reduced. Our algorithm have a smooth power management curve while approaching the performance of idea cases in temperature control and all these three metrics, comparing with the exact scheme that are highly unstable.\\
%In the simulation using real PV power generation and indoor outdoor temperature data in a typical summer day, it has been shown that our stochastic approximation algorithm has successfully tracked the dynamics of solar power variation and smart building indoor outdoor thermostats. Our results are much close to the theoretical boundary in just a few iteration while the deviates the theoretical one frequently deviates the boundary. Totally our algorithm has better performance than traditional deterministic one in all of the three goals of minimizing power loss, reduce external power consumption and achieving better user satisfaction.
\indent Actually our algorithm can be extended to a broad variety of power grid control tasks in the presence of renewable generations which are characterized by frequent fluctuations and high degree of uncertainties. Motivated by the recent trend of installing mass distributed power storage devices to cope with the variation in distributed generation, we intend to incorporate the part of power management in microgrid with distributed storage in our future work.

\end{document}